\documentclass[11pt]{aa} 

\usepackage[varg]{txfonts}
\usepackage{graphicx}
\usepackage{hyperref}
\usepackage{natbib}
\usepackage{color}
\usepackage{enumitem}

\addto\captionsenglish{\renewcommand{\abstractname}{Summary}}

\makeatletter
  \journalname{ESO Expanding Horizons Initiative}%
  \renewcommand*{\aa@manuscriptname}{White Paper Call 2025%
    \hspace{\stretch{1}}}%
  \makeatother

\begin{document}

\authorrunning{Lamprecht et al.}

\title{Resolving Galaxy Nuclei and Compact Stellar Systems \\ as Engines of Galaxy Evolution}

\author{{\fontsize{11pt}{11pt}\selectfont J. Lamprecht\inst{1,2}, I. Garland\inst{1,3}, D. Jadlovsky\inst{1}, J. Zak\inst{4} \and T. Jeřábková\inst{1}}}
\institute{
Department of Theoretical Physics and Astrophysics, Masaryk University, Kotlářská 267/2, 611 37 Brno, Czech Republic
\and
Department of Astrophysics, University of Vienna, Türkenschanzstraße 17, 1180 Vienna, Austria
\and
Department of Physics, Lancaster University, Lancaster LA1 4YB, United Kingdom
\and
Astronomical Institute of the Czech Academy of Sciences, Fri\v{c}ova 298, 25165 Ond\v{r}ejov, Czech Republic}

\date{}

\abstract{In this white paper we focus on \emph{compact stellar systems} --- star clusters, nuclear star clusters (NSCs), stripped nuclei and ultra-compact dwarfs (UCDs) --- as engines of galaxy evolution and black-hole growth. We show how the same capability also enables transformative science in active galactic nucleus (AGN) fuelling, stellar surfaces and interacting binaries, and exoplanet atmospheres. This science driver is naturally aligned with a next-generation kilometre-scale optical/IR interferometer for the 2040s era that reuses current ESO infrastructure while adding diffraction-limited integral-field spectroscopy (IFS).
}


\maketitle

\section{Galaxy evolution, compact stellar systems and black hole demographics}

{\fontsize{11pt}{12pt}\selectfont
A major outstanding challenge in galaxy formation and evolution is to understand how galactic centres grow over cosmic time, and how NSCs, black holes, and their host galaxies co-evolve \citep{Neumayer2012,kormendy2013,Neumayer2020}. The central few parsecs contain the deepest potential well, the densest stellar environments, and the clearest fossil record of a galaxy’s assembly history. Compact stellar systems trace the most intense episodes of star formation and dynamical processing. Young massive clusters, globular clusters, NSCs, stripped nuclei, and UCDs probe the highest stellar densities and most efficient channels for black hole formation. They are therefore key to understanding baryonic assembly, central mass growth, and black hole evolution. Yet even with JWST and the ELT, their internal dynamics and small-scale structures will remain unresolved beyond the Local Group. To address this, we identify three tightly connected questions:

\begin{enumerate}[label=\textbf{Q\arabic*:}]
\item \textbf{How do star clusters assemble, migrate and dissolve across cosmic time?}\\
Cluster formation efficiencies, migration in galactic potentials and tidal disruption shape the build-up of bulges, discs and stellar haloes \citep{Pfeffer2014}. Tidal tails and debris encode the dynamical history of both clusters and their host galaxies.

\item \textbf{How do dense stellar environments form black holes and exotic populations?}\\
Mass segregation, stellar collisions and binary hardening in dense cores drive the formation of stellar-mass and intermediate-mass black holes (IMBHs), as well as compact remnants and exotic stars. These processes link cluster dynamics to gravitational-wave populations and low-luminosity AGN.

\item \textbf{How do compact stellar systems build galactic centres?}\\
NSCs often coexist with central supermassive black holes \citep[SMBHs;][]{Georgiev2016,Neumayer2020}. Many scenarios invoke the inspiral and disruption of massive clusters or UCD progenitors to grow NSCs and seed black holes, but the relevant spatial scales (sub-parsec at tens of Mpc) remain beyond single-aperture telescopes. 
\end{enumerate}

\noindent A facility reaching $\sim 10$--$50\,\mu\mathrm{as}$ resolution in the optical/near-IR would resolve luminous stars and map internal proper motions in compact stellar systems out to Virgo and Fornax (16.5 and 20 Mpc away respectively) enabling:

\subsection*{{\fontsize{11pt}{11pt}\selectfont Star clusters, UCDs and high-redshift analogues}}

\begin{itemize}
\item \textbf{Dynamical IMBH mass measurements} ($10^{4}$--$10^{7}\,M_\odot$) in NSCs, UCDs and stripped nuclei, testing hypotheses that many UCDs are tidally stripped galaxy remnants \citep{Seth2014,Ahn2018};
\item \textbf{Direct tests} of mass segregation, collisional processes and binary populations in dense cluster cores;
\item \textbf{Phase-space mapping of tidal tails and debris}, constraining cluster disruption and their contribution to stellar haloes and nuclear regions;
\item \textbf{Connecting LRDs to local systems.} Very massive proto-UCDs with top-heavy initial mass functions (IMFs) may appear as compact, quasar-like objects \citep{Jerabkova2017, Kroupa2020}. JWST has revealed extremely compact, red ``Little Red Dots'' (LRDs), interpreted as rapidly accreting BH seeds \citep{Bellovary2025}. Studying nearby UCDs, stripped nuclei and NSCs with micro-arcsecond astrometry links these local laboratories to their high-redshift analogues.

\end{itemize}

\subsection*{{\fontsize{11pt}{11pt}\selectfont Nuclear star clusters and the low-mass BH regime}}

\begin{itemize}
    \item \textbf{Measuring black holes in the low-mass regime.} NSCs provide unique leverage on BH demographics at $10^{4}$--$10^{7}\,M_\odot$ \citep{Pechetti2020}, where $M_{\rm BH}$–$M_\star$ relations are weakest \citep{ReinesVolonteri2015,kormendy2013} and seeding models diverge. A micro-arcsecond facility would yield precise dynamical BH masses in dwarf galaxies, late-type spirals and bulgeless discs;
    \item \textbf{Resolving the orbital make-up of NSCs and dense clusters} out to several Mpc, allowing both proper motions and line-of-sight velocities to be measured directly. Linking these resolved stars to their underlying stellar populations would provide direct constraints on the formation and growth of galactic centres, revealing whether NSCs are assembled predominantly through in-situ star formation, cluster infall, or episodic accretion events, while mapping anisotropy, rotation and disc-like components.
\end{itemize}

\noindent Combining such dynamical maps with ELT-class IFS will enable chemo-dynamical modelling of NSCs and their connection to host-galaxy scaling relations, closing the gap between stellar-mass and supermassive black holes.

\section{{\fontsize{11pt}{11pt}\selectfont Additional discovery space in the same regime}}

The same angular resolution and astrometric precision that compact stellar systems require naturally enable breakthroughs in other fields where structure is confined to sub-parsec or sub-stellar-radius scales.

\subsection*{{\fontsize{11pt}{11pt}\selectfont Secular AGN triggering and fuelling}}

Connections between SMBHs and host-galaxy properties (e.g., $M_{\mathrm{BH}}$--$\sigma$, $M_{\mathrm{BH}}$--bulge mass) indicate co-evolution \citep{haring2004,simmons2017}. While mergers account for only 15–35\% of BH growth since $z\sim3$ \citep{martin2018,mcalpine2020}, the dominant fuelling mechanisms remain poorly constrained. Disc-dominated galaxies offer a merger-free environment to study secular BH growth, as they have been merger-free since $z\sim2$ \citep{martig2012}. Large-scale bars are present in $\sim$30\% of discs, and AGN fractions are higher in strongly barred systems \citep[e.g.][]{galloway2015, silvalima2022, garland2024}. Whether bars directly trigger or fuel AGN, however, remains unclear.
A micro-arcsecond ELT/VLTI-era facility would enable:
\begin{itemize}
\item \textbf{Tracing gas flow in AGN hosts}. High-resolution morphology and optical IFS would map gas kinematics and ionisation across entire galaxies, distinguishing star-formation from AGN excitation, and identifying gas inflows or outflows. We could directly test how bars and spiral arms regulate these flows;

    \item \textbf{Resolving the kiloparsec problem}. There are many mechanisms by which gas can travel to the central kiloparsec of galaxies, but how it then gets to the central SMBH is debated \citep{goodman2003, hopkins2011}. Simulations provide us with some insight, but there is still a need for observational confirmation. With a resolution of With a resolution of $\sim 100\,\mu\mathrm{as}$, we can resolve $0.1\,\mathrm{pc}$ in AGN up to redshift $z\sim0.05$, potentially providing us with observational confirmation with a large sample of AGN;

    \item \textbf{Alternative fuelling mechanisms}. With such precise and accurate morphology, we can observe AGN in disc galaxies that appear to lack both a bar and a bulge. Such systems suggest that nuclear activity may occur without mergers or large-scale stellar structures. Enhanced imaging may reveal previously unresolved bars or bulges, or confirm that no such features exist. In the latter case, the results would point to additional, yet unrecognised pathways for AGN triggering and fuelling, motivating revised models of secular black-hole growth.
\end{itemize}

\noindent With such a facility, we could make phenomenal strides in the field of AGN feedback, fuelling and galaxy co-evolution. This would transform our understanding of merger-free black-hole growth.

\subsection*{{\fontsize{11pt}{11pt}\selectfont Stellar surfaces, mass loss and astrometric orbits for distant systems}}

Stellar evolution and feedback depend sensitively on surface convection, pulsation, mass loss, and transfer in the case of interacting binaries. Current interferometers have begun to resolve the surfaces of evolved stars and measure asymmetric winds and hotspots. A next-generation facility would extend this to:

\begin{itemize}
\item \textbf{Resolving stellar surfaces}. With the current maximum angular resolution of $\sim 0.8\,\mathrm{mas}$ ($H$-band with PIONIER at 200-m baseline), interferometric imaging is limited mostly to giants and supergiants within $\lesssim 2\,\mathrm{kpc}$. Existing observations already reveal convection cells and other surface structures \citep[e.g.,][]{paladini18, rosalez24}. Increasing the angular resolution to $\sim 100\,\mu\mathrm{as}$ would broaden the accessible stellar sample, resolve finer structures, and combined with AO-driven sensitivity gains, allow imaging of more distant and faint objects, including circumstellar environments of evolved supergiants in the Galactic Centre and the Magellanic Clouds;
\item \textbf{Binary interaction}. Higher resolution would enable detailed imaging of circumstellar and circumbinary structures, including discs 
of Be stars and B[e] supergiants \citep[e.g.,]{wheelwright12, hofmann22} and envelopes shaped by unseen companions \citep[e.g.,][]{ohnaka24}. It would also resolve Roche-lobe–filling yellow and red (super)giants \citep[e.g.,][]{wittkowski17, merc24}. Both the number of accessible interacting binaries and the proximity to their surface structures would increase dramatically;
\item \textbf{Astrometric orbits}. VLTI already detects close stellar and sub-stellar companions with sub-mas separations \citep{galenne23}, probing parameter space inaccessible to most instruments. Enhanced baselines and sensitivity would reveal even closer, fainter companions and improve astrometric precision, yielding more accurate orbits and dynamical masses.

\end{itemize}

\noindent These observations feed directly into stellar-evolution and population-synthesis models that are fundamental to cluster and galaxy evolution studies.

\subsection*{{\fontsize{11pt}{11pt}\selectfont Exoplanet atmospheres as compact thermal structures}}

ESO facilities (HARPS, ESPRESSO, VLT/ELT imagers), together with space missions (JWST and the future Ariel and HWO), are transforming exoplanet science by providing precise masses \citep{hobson}, detections of non-transiting companions \citep{gonzalez}, and insights into orbital architectures \citep{zak}. ELT instrumentation will unite extreme radial velocity precision with high-contrast imaging \citep{palle}. JWST now delivers molecular abundances, cloud properties, and evidence for time-variable atmospheric structure \citep{rusta, edwards}. Yet all current observations remain spatially unresolved, limiting our understanding of circulation, cloud patterns, and surface inhomogeneities. GRAVITY+ provides exquisite astrometry and orbital motion, while direct imaging constrains spin rates and atmospheric properties of young giants—but only for bright, widely separated systems under present baseline limits. A next-generation, kilometre-scale interferometer would:
\begin{itemize}
\item \textbf{Resolve atmospheric structures and circulation.} By spatially mapping clouds, hotspots, atmospheric bands, and weather systems on nearby brown dwarfs and giant exoplanets, and through time-series imaging that traces evolving circulation patterns, these observations will provide direct constraints for calibrating atmospheric models \citep{showman}. Additionally, these resolved maps would provide essential context for interpreting disc-integrated spectra from upcoming missions such as Ariel \citep{tin18} and NASA’s HWO \citep{hwo}, enabling three-dimensional constraints on atmospheric variability \citep{challener}.
\item \textbf{Detect exomoons, circumplanetary discs, and probe planetary dynamics.} Resolve planet–moon separations around nearby giants to obtain the first unambiguous exomoon detections, dynamical masses, and constraints complementary to JWST searches \citep{kipping22, kipping25}. The same spatial precision would allow direct measurements of planetary oblateness and rotation rates \citep{cassese}, probing internal structure and angular-momentum evolution.
\item \textbf{Access older, cooler and close-in giant planets.} Observe nearby systems too close for GRAVITY+ and non-transiting for JWST, enabling spatially resolved or disc-integrated spectroscopy. These observations would test formation pathways for mature giant planets and substellar objects \citep{hatzes, gilbert} and extend atmospheric characterisation beyond young, self-luminous systems.
\end{itemize}

\noindent No other foreseeable technique offers comparable spatial resolution or scientific reach. Kilometre-scale optical/IR interferometry would thus be uniquely transformative for exoplanet characterisation.

\section{Facility requirements}

The above science consistently points to the need for:

\begin{itemize}
\item Angular resolution of $\sim 10$--$50\,\mu\mathrm{as}$ at optical/NIR wavelengths;
\item Astrometric precision at the $\sim$10~$\mu$as level over multi-year baselines;
\item Sensitivity to detect evolved stars (faint stellar targets) in clusters and NSCs at Virgo/Fornax distances while simultaneously coping with very bright nearby sources/ galactic nuclei
\item Dynamic range sufficient for compact structures around bright sources (AGN, stellar photospheres, exoplanets);
\item Synergy with ELT-class IFS for chemo-dynamics, and with high-energy and gravitational-wave facilities.
\end{itemize}

\noindent This white paper builds on the interferometric vision outlined by \cite{bourdarot2024}, who present a kilometre-scale optical/IR array capable of delivering sub-milliarcsecond imaging. In our proposed configuration, we retain the core elements of the optical/IR array design but introduce an additional, complementary pathway: equipping one of the central telescopes (i.e., ``X1'') with a diffraction-limited IFS, conceptually similar to HARMONI on the ELT. Importantly, the IFS and the interferometric array need not operate simultaneously. X1 could function as a standalone diffraction-limited telescope for spectroscopy whenever the rest of the array is used for interferometry, and vice versa. This separation also avoids the substantial technical challenges of combining interferometric and IFS modes in the same optical train. A repurposed or recycled ELT IFS module (e.g. a HARMONI derivative) would minimise cost, reduce development risk, and take advantage of ESO’s existing hardware ecosystem. This would greatly increase the community's access to diffraction-limited spectroscopy, which is constrained by the limited availability and oversubscription of existing IFS facilities. To ensure scientific flexibility, the new X-telescopes should operate independently and provide both long and short baselines. Extreme resolution is valuable, but without shorter baselines many extended sources would be over-resolved. A compact configuration --- analogous to the role of the ACA 7 m array and single-dish data in ALMA --- would allow the facility to probe a continuous range of spatial scales. This could be achieved by reusing existing UT positions (if available) or by constructing a new set of closer X-stations. Because the long-term scheduling of the UTs is uncertain, the system must remain fully functional without UT participation. A network of 3 -- 4 dedicated X-telescopes would already provide a robust standalone interferometer with sufficient baseline coverage for imaging. Overall, the architecture combines long baselines with AO-assisted, diffraction-limited spectroscopy to enable detailed studies of galactic nuclei, AGN environments, compact stellar systems, stellar evolution, and exoplanets, while building on ESO’s established interferometric expertise.

\section{Sustainability}

{\fontsize{11pt}{12pt}\selectfont
Any next-generation interferometer must be designed with long-term sustainability in mind. A central principle is to build on existing ESO infrastructure, avoiding new large domes, heavy foundations, or additional energy-intensive cooling systems. By reusing delay lines, tunnels, operational buildings, and parts of the Paranal technical ecosystem, the construction-related environmental footprint is significantly reduced. A modular array layout further enables phased deployment, gradual upgrades, and smaller-aperture telescopes that require less material and maintenance.
Renewable-energy integration offers a realistic path toward low-carbon operations. The Paranal–Armazones region has exceptionally high solar irradiance, and ESO has already demonstrated large-scale feasibility through the 9 MW solar farm installed in 2022, which will supply power to the ELT. Extending similar solar-hybrid micro-grids -- combining photovoltaics, battery storage, and existing diesel backup -- to new interferometric stations on extended baselines provides stable, low-carbon power for delay lines, AO, metrology, and cryogenic subsystems without major retrofits to central VLT services. Supplementary use of existing VISTA or VLT infrastructure further limits new construction.
Operational sustainability is supported by modern AO, automated calibrations, and machine-learning–assisted pipelines, all improving instrument efficiency and stability. Reusing major subsystems (e.g., repurposing a HARMONI-derived IFS once the ELT instrumentation suite evolves) promotes a circular hardware life cycle within ESO. Overall, leveraging existing expertise and infrastructure while integrating emerging ELT technologies maximizes scientific return per environmental cost. The approach emphasizes technical longevity, realistic energy-efficiency improvements, renewable-energy use, and component reusability, aligning with ESO’s broader commitment to sustainable observatory development.

}}

\bibliographystyle{aa}   
\bibliography{References} 

@ARTICLE{Kroupa2020,
       author = {{Kroupa}, Pavel and {Subr}, Ladislav and {Jerabkova}, Tereza and {Wang}, Long},
        title = "{Very high redshift quasars and the rapid emergence of supermassive black holes}",
      journal = {\mnras},
         year = 2020,
        month = nov,
       volume = {498},
       number = {4},
        pages = {5652-5683},
          doi = {10.1093/mnras/staa2276},
archivePrefix = {arXiv},
       eprint = {2007.14402},
 primaryClass = {astro-ph.GA},
       adsurl = {https://ui.adsabs.harvard.edu/abs/2020MNRAS.498.5652K}
}

@ARTICLE{Neumayer2012,
       author = {{Neumayer}, Nadine and {Walcher}, C. Jakob},
        title = "{Are Nuclear Star Clusters the Precursors of Massive Black Holes?}",
      journal = {Advances in Astronomy},
         year = 2012,
        month = jan,
       volume = {2012},
          eid = {709038},
        pages = {709038},
          doi = {10.1155/2012/709038},
archivePrefix = {arXiv},
       eprint = {1201.4950},
 primaryClass = {astro-ph.CO},
       adsurl = {https://ui.adsabs.harvard.edu/abs/2012AdAst2012E..15N}
}

@ARTICLE{Pechetti2020,
       author = {{Pechetti}, Renuka and {Seth}, Anil and {Neumayer}, Nadine and {Georgiev}, Iskren and {Kacharov}, Nikolay and {den Brok}, Mark},
        title = "{Luminosity Models and Density Profiles for Nuclear Star Clusters for a Nearby Volume-limited Sample of 29 Galaxies}",
      journal = {\apj},
         year = 2020,
        month = sep,
       volume = {900},
       number = {1},
          eid = {32},
        pages = {32},
          doi = {10.3847/1538-4357/abaaa7},
archivePrefix = {arXiv},
       eprint = {1911.09686},
 primaryClass = {astro-ph.GA},
       adsurl = {https://ui.adsabs.harvard.edu/abs/2020ApJ...900...32P}
}

@ARTICLE{Ahn2018,
       author = {{Ahn}, Christopher P. and {Seth}, Anil C. and {Cappellari}, Michele and {Krajnovi{\'c}}, Davor and {Strader}, Jay and {Voggel}, Karina T. and {Walsh}, Jonelle L. and {Bahramian}, Arash and {Baumgardt}, Holger and {Brodie}, Jean and et al.},
        title = "{The Black Hole in the Most Massive Ultracompact Dwarf Galaxy M59-UCD3}",
      journal = {\apj},
         year = 2018,
        month = may,
       volume = {858},
       number = {2},
          eid = {102},
        pages = {102},
          doi = {10.3847/1538-4357/aabc57},
archivePrefix = {arXiv},
       eprint = {1804.02399},
 primaryClass = {astro-ph.GA},
       adsurl = {https://ui.adsabs.harvard.edu/abs/2018ApJ...858..102A}
}

@ARTICLE{Seth2014,
       author = {{Seth}, Anil C. and {van den Bosch}, Remco and {Mieske}, Steffen and {Baumgardt}, Holger and {Brok}, Mark Den and {Strader}, Jay and {Neumayer}, Nadine and {Chilingarian}, Igor and {Hilker}, Michael and {McDermid}, Richard and et al.},
        title = "{A supermassive black hole in an ultra-compact dwarf galaxy}",
      journal = {\nat},
         year = 2014,
        month = sep,
       volume = {513},
       number = {7518},
        pages = {398-400},
          doi = {10.1038/nature13762},
archivePrefix = {arXiv},
       eprint = {1409.4769},
 primaryClass = {astro-ph.GA},
       adsurl = {https://ui.adsabs.harvard.edu/abs/2014Natur.513..398S}
}

@ARTICLE{Georgiev2016,
       author = {{Georgiev}, Iskren Y. and {B{\"o}ker}, Torsten and {Leigh}, Nathan and {L{\"u}tzgendorf}, Nora and {Neumayer}, Nadine},
        title = "{Masses and scaling relations for nuclear star clusters, and their co-existence with central black holes}",
      journal = {\mnras},
         year = 2016,
        month = apr,
       volume = {457},
       number = {2},
        pages = {2122-2138},
          doi = {10.1093/mnras/stw093},
archivePrefix = {arXiv},
       eprint = {1601.02613},
 primaryClass = {astro-ph.GA},
       adsurl = {https://ui.adsabs.harvard.edu/abs/2016MNRAS.457.2122G}
}

@ARTICLE{Pfeffer2014,
       author = {{Pfeffer}, J. and {Griffen}, B.~F. and {Baumgardt}, H. and {Hilker}, M.},
        title = "{Contribution of stripped nuclear clusters to globular cluster and ultracompact dwarf galaxy populations}",
      journal = {\mnras},
         year = 2014,
        month = nov,
       volume = {444},
       number = {4},
        pages = {3670-3683},
          doi = {10.1093/mnras/stu1705},
archivePrefix = {arXiv},
       eprint = {1408.4467},
 primaryClass = {astro-ph.GA},
       adsurl = {https://ui.adsabs.harvard.edu/abs/2014MNRAS.444.3670P}
}

@ARTICLE{Neumayer2020,
       author = {{Neumayer}, Nadine and {Seth}, Anil and {B{\"o}ker}, Torsten},
        title = "{Nuclear star clusters}",
      journal = {\aapr},
         year = 2020,
        month = jul,
       volume = {28},
       number = {1},
          eid = {4},
        pages = {4},
          doi = {10.1007/s00159-020-00125-0},
archivePrefix = {arXiv},
       eprint = {2001.03626},
 primaryClass = {astro-ph.GA},
       adsurl = {https://ui.adsabs.harvard.edu/abs/2020A&ARv..28....4N}
}

@ARTICLE{Jerabkova2017,
       author = {{Je{\v{r}}{\'a}bkov{\'a}}, T. and {Kroupa}, P. and {Dabringhausen}, J. and {Hilker}, M. and {Bekki}, K.},
        title = "{The formation of ultra compact dwarf galaxies and massive globular clusters. Quasar-like objects to test for a variable stellar initial mass function}",
      journal = {\aap},
         year = 2017,
        month = dec,
       volume = {608},
          eid = {A53},
        pages = {A53},
          doi = {10.1051/0004-6361/201731240},
archivePrefix = {arXiv},
       eprint = {1708.07127},
 primaryClass = {astro-ph.GA},
       adsurl = {https://ui.adsabs.harvard.edu/abs/2017A&A...608A..53J}
}

@ARTICLE{Bellovary2025,
       author = {{Bellovary}, Jillian},
        title = "{Little Red Dots Are Tidal Disruption Events in Runaway-collapsing Clusters}",
      journal = {\apjl},
         year = 2025,
        month = may,
       volume = {984},
       number = {2},
          eid = {L55},
        pages = {L55},
          doi = {10.3847/2041-8213/adce6c},
archivePrefix = {arXiv},
       eprint = {2501.03309},
 primaryClass = {astro-ph.GA},
       adsurl = {https://ui.adsabs.harvard.edu/abs/2025ApJ...984L..55B}
}

@article{bourdarot2024,
      title={Kilometer-baseline interferometry: science drivers for the next generation instrument}, 
      author={G. Bourdarot and F. Eisenhauer},
      year={2024},
      eprint={2410.22063},
      archivePrefix={arXiv},
      primaryClass={astro-ph.IM},
      url={https://arxiv.org/abs/2410.22063}, 
}

@ARTICLE{galloway2015,
    author = {{Galloway}, M.~A. and {Willett}, K.~W. and {Fortson}, L.~F. and {Cardamone}, C.~N. and {Schawinski}, K. and {Cheung}, E. and {Lintott}, C.~J. and {Masters}, K.~L. and {Melvin}, T. and {Simmons}, B.~D.},
    title = "{Galaxy Zoo: the effect of bar-driven fuelling on the presence of an active galactic nucleus in disc galaxies}",
    journal = {\mnras},
    archivePrefix = "arXiv",
    eprint = {1502.01033},
    year = 2015,
    month = apr,
    volume = 448,
    pages = {3442-3454},
    doi = {10.1093/mnras/stv235},
    adsurl = {http://adsabs.harvard.edu/abs/2015MNRAS.448.3442G}}

@ARTICLE{garland2024,
    author = {{Garland}, Izzy L. and {Walmsley}, Mike and {Silcock}, Maddie S. and {Potts}, Leah M. and {Smith}, Josh and {Simmons}, Brooke D. and {Lintott}, Chris J. and {Smethurst}, Rebecca J. and {Dawson}, James M. and {Keel}, William C. and {Kruk}, Sandor and {Mantha}, Kameswara Bharadwaj and {Masters}, Karen L. and {O'Ryan}, David and {Popp}, J{\"u}rgen J. and {Thorne}, Matthew R.},
    title = "{Galaxy Zoo DESI: large-scale bars as a secular mechanism for triggering AGNs}",
    journal = {\mnras},
    year = 2024,
    month = aug,
    volume = {532},
    number = {2},
    pages = {2320-2330},
    doi = {10.1093/mnras/stae1620},
    archivePrefix = {arXiv},
    eprint = {2406.20096},
    primaryClass = {astro-ph.GA},
    adsurl = {https://ui.adsabs.harvard.edu/abs/2024MNRAS.532.2320G}}

@ARTICLE{goodman2003,
    author = {{Goodman}, Jeremy},
    title = "{Self-gravity and quasi-stellar object discs}",
    journal = {\mnras},
    year = 2003,
    month = mar,
    volume = {339},
    number = {4},
    pages = {937-948},
    doi = {10.1046/j.1365-8711.2003.06241.x},
    archivePrefix = {arXiv},
    eprint = {astro-ph/0201001},
    primaryClass = {astro-ph},
    adsurl = {https://ui.adsabs.harvard.edu/abs/2003MNRAS.339..937G}}

@ARTICLE{haring2004,
    author = {{H{\"a}ring}, Nadine and {Rix}, Hans-Walter},
    title = "{On the Black Hole Mass-Bulge Mass Relation}",
    journal = {\apj},
    year = "2004",
    month = "Apr",
    volume = {604},
    number = {2},
    pages = {L89-L92},
    doi = {10.1086/383567},
    archivePrefix = {arXiv},
    eprint = {astro-ph/0402376},
    primaryClass = {astro-ph},
    adsurl = {https://ui.adsabs.harvard.edu/abs/2004ApJ...604L..89H}}

@ARTICLE{hopkins2011,
    author = {{Hopkins}, Philip F. and {Quataert}, Eliot},
    title = "{An analytic model of angular momentum transport by gravitational torques: from galaxies to massive black holes}",
    journal = {\mnras},
    year = 2011,
    month = aug,
    volume = {415},
    number = {2},
    pages = {1027-1050},
    doi = {10.1111/j.1365-2966.2011.18542.x},
    archivePrefix = {arXiv},
    eprint = {1007.2647},
    primaryClass = {astro-ph.CO},
    adsurl = {https://ui.adsabs.harvard.edu/abs/2011MNRAS.415.1027H}}

@ARTICLE{ReinesVolonteri2015,
       author = {{Reines}, Amy E. and {Volonteri}, Marta},
        title = "{Relations between Central Black Hole Mass and Total Galaxy Stellar Mass in the Local Universe}",
      journal = {\apj},
         year = 2015,
        month = nov,
       volume = {813},
       number = {2},
          eid = {82},
        pages = {82},
          doi = {10.1088/0004-637X/813/2/82},
archivePrefix = {arXiv},
       eprint = {1508.06274},
 primaryClass = {astro-ph.GA},
       adsurl = {https://ui.adsabs.harvard.edu/abs/2015ApJ...813...82R}
}

@ARTICLE{kormendy2013,
    title = {Coevolution (or not) of supermassive black holes and host galaxies},
    author = {Kormendy, J. and Ho, L. C.},
    journal = {\araa},
    volume = {51},
    pages = {511-653},
    year = {2013},
    publisher = {Annual Reviews}}

@ARTICLE{martig2012,
    author = {{Martig}, Marie and {Bournaud}, Fr{\'e}d{\'e}ric and {Croton}, Darren J. and {Dekel}, Avishai and {Teyssier}, Romain},
    title = "{A Diversity of Progenitors and Histories for Isolated Spiral Galaxies}",
    journal = {\apj},
    year = 2012,
    month = sep,
    volume = {756},
    number = {1},
    eid={26},
    pages = {26},
    doi = {10.1088/0004-637X/756/1/26},
    archivePrefix = {arXiv},
    eprint = {1201.1079},
    primaryClass = {astro-ph.CO},
    adsurl = {https://ui.adsabs.harvard.edu/abs/2012ApJ...756...26M}}

@ARTICLE{martin2018,
    author = {{Martin}, G. and {Kaviraj}, S. and {Volonteri}, M. and {Simmons}, B.~D. and {Devriendt}, J.~E.~G. and {Lintott}, C.~J. and {Smethurst}, R.~J. and {Dubois}, Y. and {Pichon}, C.},
    title = "{Normal black holes in bulge-less galaxies: the largely quiescent, merger-free growth of black holes over cosmic time}",
    journal = {\mnras},
    year = 2018,
    month = may,
    volume = {476},
    number = {2},
    pages = {2801-2812},
    doi = {10.1093/mnras/sty324},
    archivePrefix = {arXiv},
    eprint = {1801.09699},
    primaryClass = {astro-ph.GA},
    adsurl = {https://ui.adsabs.harvard.edu/abs/2018MNRAS.476.2801M}}

@ARTICLE{mcalpine2020,
    author = {{McAlpine}, Stuart and {Harrison}, Chris M. and {Rosario}, David J. and {Alexander}, David M. and {Ellison}, Sara L. and {Johansson}, Peter H. and {Patton}, David R.},
    title = "{Galaxy mergers in EAGLE do not induce a significant amount of black hole growth yet do increase the rate of luminous AGN}",
    journal = {\mnras},
    year = 2020,
    month = jun,
    volume = {494},
    number = {4},
    pages = {5713-5733},
    doi = {10.1093/mnras/staa1123},
    archivePrefix = {arXiv},
    eprint = {2002.00959},
    primaryClass = {astro-ph.GA},
    adsurl = {https://ui.adsabs.harvard.edu/abs/2020MNRAS.494.5713M}}

@ARTICLE{silvalima2022,
    author = {{Silva-Lima}, Luiz A. and {Martins}, Lucimara P. and {Coelho}, Paula R.~T. and {Gadotti}, Dimitri A.},
    title = "{Revisiting the role of bars in AGN fuelling with propensity score sample matching}",
    journal = {\aap},
    year = 2022,
    month = may,
    volume = {661},
    eid = {A105},
    pages = {A105},
    doi = {10.1051/0004-6361/202142432},    
    archivePrefix = {arXiv},
    eprint = {2203.07794},
    primaryClass = {astro-ph.GA},
    adsurl = {https://ui.adsabs.harvard.edu/abs/2022A&A...661A.105S}}

@ARTICLE{simmons2017,
    author = {{Simmons}, B.~D. and {Smethurst}, R.~J. and {Lintott}, C.},
    title = "{Supermassive black holes in disc-dominated galaxies outgrow their bulges and co-evolve with their host galaxies}",
    journal = {\mnras},
    year = 2017,
    month = sep,
    volume = {470},
    number = {2},
    pages = {1559-1569},
    doi = {10.1093/mnras/stx1340},
    archivePrefix = {arXiv},
    eprint = {1705.10793},
    primaryClass = {astro-ph.GA},
    adsurl = {https://ui.adsabs.harvard.edu/abs/2017MNRAS.470.1559S}}

@ARTICLE{galenne23,
       author = {{Gallenne}, A. and {M{\'e}rand}, A. and {Kervella}, P. and {Graczyk}, D. and {Pietrzy{\'n}ski}, G. and {Gieren}, W. and {Pilecki}, B.},
        title = "{The Araucaria project: High-precision orbital parallaxes and masses of binary stars. I. VLTI/GRAVITY observations of ten double-lined spectroscopic binaries}",
      journal = {\aap},
         year = 2023,
        month = apr,
       volume = {672},
          eid = {A119},
        pages = {A119},
          doi = {10.1051/0004-6361/202245712},
archivePrefix = {arXiv},
       eprint = {2302.12960},
 primaryClass = {astro-ph.SR},
       adsurl = {https://ui.adsabs.harvard.edu/abs/2023A&A...672A.119G}
}

@ARTICLE{wheelwright12,
       author = {{Wheelwright}, H.~E. and {de Wit}, W.~J. and {Oudmaijer}, R.~D. and {Vink}, J.~S.},
        title = "{VLTI/AMBER observations of the binary B[e] supergiant HD 327083}",
      journal = {\aap},
         year = 2012,
        month = feb,
       volume = {538},
          eid = {A6},
        pages = {A6},
          doi = {10.1051/0004-6361/201117766},
archivePrefix = {arXiv},
       eprint = {1201.2866},
 primaryClass = {astro-ph.SR},
       adsurl = {https://ui.adsabs.harvard.edu/abs/2012A&A...538A...6W}
}

@ARTICLE{hofmann22,
       author = {{Hofmann}, K.-H. and {Bensberg}, A. and {Schertl}, D. and {Weigelt}, G. and {Wolf}, S. and {Meilland}, A. and {Millour}, F. and {Waters}, L.~B.~F.~M. and {Kraus}, S. and {Ohnaka}, K. and {Lopez}, B. and {Petrov}, R.~G. and {Lagarde}, S. and {Berio}, Ph. and {Allouche}, F. and {Robbe-Dubois}, S. and {Jaffe}, W. and {Henning}, Th. and {Paladini}, C. and {Sch{\"o}ller}, M. and {M{\'e}rand}, A. and {Glindemann}, A. and {Beckmann}, U. and {Heininger}, M. and {Bettonvil}, F. and {Zins}, G. and {Woillez}, J. and {Bristow}, P. and {Stee}, P. and {Vakili}, F. and {van Boekel}, R. and {Hogerheijde}, M.~R. and {Dominik}, C. and {Augereau}, J.-C. and {Matter}, A. and {Hron}, J. and {Pantin}, E. and {Rivinius}, Th. and {de Wit}, W.-J. and {Varga}, J. and {Klarmann}, L. and {Meisenheimer}, K. and {G{\'a}mez Rosas}, V. and {Burtscher}, L. and {Leftley}, J. and {Isbell}, J.~W. and {Yoffe}, G. and {Kokoulina}, E. and {Danchi}, W.~C. and {Cruzal{\`e}bes}, P. and {Domiciano de Souza}, A. and {Drevon}, J. and {Hocd{\'e}}, V. and {Kreplin}, A. and {Labadie}, L. and {Connot}, C. and {Nu{\ss}baum}, E. and {Lehmitz}, M. and {Antonelli}, P. and {Graser}, U. and {Leinert}, C.},
        title = "{VLTI-MATISSE L- and N-band aperture-synthesis imaging of the unclassified B[e] star FS Canis Majoris}",
      journal = {\aap},
         year = 2022,
        month = feb,
       volume = {658},
          eid = {A81},
        pages = {A81},
          doi = {10.1051/0004-6361/202141601},
archivePrefix = {arXiv},
       eprint = {2111.12458},
 primaryClass = {astro-ph.SR},
       adsurl = {https://ui.adsabs.harvard.edu/abs/2022A&A...658A..81H}
}

@ARTICLE{rosalez24,
       author = {{Rosales-Guzm{\'a}n}, A. and {Sanchez-Bermudez}, J. and {Paladini}, C. and {Freytag}, B. and {Wittkowski}, M. and {Alberdi}, A. and {Baron}, F. and {Berger}, J.-P. and {Chiavassa}, A. and {H{\"o}fner}, S. and {Jorissen}, A. and {Kervella}, P. and {Le Bouquin}, J.-B. and {Marigo}, P. and {Montarg{\`e}s}, M. and {Trabucchi}, M. and {Tsvetkova}, S. and {Sch{\"o}del}, R. and {Van Eck}, S.},
        title = "{A new dimension in the variability of AGB stars: Convection patterns size changes with pulsation}",
      journal = {\aap},
         year = 2024,
        month = aug,
       volume = {688},
          eid = {A124},
        pages = {A124},
          doi = {10.1051/0004-6361/202349112},
archivePrefix = {arXiv},
       eprint = {2405.10164},
 primaryClass = {astro-ph.SR},
       adsurl = {https://ui.adsabs.harvard.edu/abs/2024A&A...688A.124R}
}

@ARTICLE{paladini18,
       author = {{Paladini}, C. and {Baron}, F. and {Jorissen}, A. and {Le Bouquin}, J.-B. and {Freytag}, B. and {van Eck}, S. and {Wittkowski}, M. and {Hron}, J. and {Chiavassa}, A. and {Berger}, J.-P. and {Siopis}, C. and {Mayer}, A. and {Sadowski}, G. and {Kravchenko}, K. and {Shetye}, S. and {Kerschbaum}, F. and {Kluska}, J. and {Ramstedt}, S.},
        title = "{Large granulation cells on the surface of the giant star {\ensuremath{\pi}}$^{1}$ Gruis}",
      journal = {\nat},
         year = 2018,
        month = jan,
       volume = {553},
       number = {7688},
        pages = {310-312},
          doi = {10.1038/nature25001},
       adsurl = {https://ui.adsabs.harvard.edu/abs/2018Natur.553..310P}
}

@ARTICLE{ohnaka24,
       author = {{Ohnaka}, K. and {Hofmann}, K.-H. and {Weigelt}, G. and {van Loon}, J. Th. and {Schertl}, D. and {Goldman}, S.~R.},
        title = "{Imaging the innermost circumstellar environment of the red supergiant WOH G64 in the Large Magellanic Cloud}",
      journal = {\aap},
         year = 2024,
        month = nov,
       volume = {691},
          eid = {L15},
        pages = {L15},
          doi = {10.1051/0004-6361/202451820},
archivePrefix = {arXiv},
       eprint = {2412.01921},
 primaryClass = {astro-ph.SR},
       adsurl = {https://ui.adsabs.harvard.edu/abs/2024A&A...691L..15O}
}

@ARTICLE{wittkowski17,
       author = {{Wittkowski}, M. and {Abell{\'a}n}, F.~J. and {Arroyo-Torres}, B. and {Chiavassa}, A. and {Guirado}, J.~C. and {Marcaide}, J.~M. and {Alberdi}, A. and {de Wit}, W.~J. and {Hofmann}, K.-H. and {Meilland}, A. and {Millour}, F. and {Mohamed}, S. and {Sanchez-Bermudez}, J.},
        title = "{Multi-epoch VLTI-PIONIER imaging of the supergiant V766 Cen}",
      journal = {\aap},
         year = 2017,
        month = sep,
       volume = {606},
          eid = {L1},
        pages = {L1},
          doi = {10.1051/0004-6361/201731569},
archivePrefix = {arXiv},
       eprint = {1709.09430},
 primaryClass = {astro-ph.SR},
       adsurl = {https://ui.adsabs.harvard.edu/abs/2017A&A...606L...1W}
}

@ARTICLE{merc24,
       author = {{Merc}, Jaroslav and {Boffin}, Henri M.~J.},
        title = "{Unequivocal detection of the tidal deformation of a red giant in a binary system via interferometry}",
      journal = {\aap},
         year = 2024,
        month = dec,
       volume = {692},
          eid = {A218},
        pages = {A218},
          doi = {10.1051/0004-6361/202451952},
archivePrefix = {arXiv},
       eprint = {2411.14621},
 primaryClass = {astro-ph.SR},
       adsurl = {https://ui.adsabs.harvard.edu/abs/2024A&A...692A.218M}
}

\end{document}